\documentclass[fleqn,usenatbib]{mnras}

\usepackage{graphicx}	% Including figure files
\usepackage{amsmath}	% Advanced maths commands
\usepackage{amssymb}
\usepackage[T1]{fontenc}
\usepackage{etoolbox}
\robustify{\cite}

\usepackage{newtxtext,newtxmath}

\title[Evolution of reflection spectra]{Evolution of accretion disc reflection spectra due to a Type I X-ray burst}
\graphicspath{{./}{figuresPaper/}}
\author[J. Speicher et al.]{
J. Speicher,$^{1}$\thanks{E-mail: jspeicher3@gatech.edu}
D. R. Ballantyne$^{1}$
and P. C. Fragile$^{2}$
\\
$^{1}$Center for Relativistic Astrophysics, School of Physics, Georgia
  Institute of Technology, 837 State Street, Atlanta, GA 30332-0430, USA\\
$^{2}$Department of Physics \& Astronomy, College of Charleston, 66 George St., Charleston, SC 29424, USA
}

\date{Accepted XXX. Received YYY; in original form ZZZ}

\pubyear{2021}

\begin{document}
\label{firstpage}
\pagerange{\pageref{firstpage}--\pageref{lastpage}}
\maketitle

\begin{abstract}
Irradiation of the accretion disc causes reflection signatures in the observed X-ray spectrum, encoding important information about the disc structure and density. A Type I X-ray burst will strongly irradiate the accretion disc and alter its properties. Previous numerical simulations predicted the evolution of the accretion disc due to an X-ray burst. Here, we process time-averaged simulation data of six time intervals to track changes in the reflection spectrum from the burst onset to just past its peak. We divide the reflecting region of the disc within $r\lesssim50$ km into 6-7 radial zones for every time interval and compute the reflection spectra for each zone. We integrate these reflection spectra to obtain a total reflection spectrum per time interval. The burst ionizes and heats the disc, which gradually weakens all emission lines. Compton scattering and bremsstrahlung rates increase in the disc during the burst rise, and the soft excess at $<$3 keV rises from $\approx4$\% to $\approx38$\% of the total emission at the burst peak.
A soft excess is expected to be ubiquitous in the reflection spectra of X-ray bursts. Structural disc changes such as inflation because of heating or drainage of the inner disc due to Poynting-Robertson drag affect the strength of the soft excess. %Geometric changes of the accretion disc%, such as the retreat of the inner edge of the disc due to Poynting- Robertson drag and disc inflation, discourage and promote disc heating respectively and therefore increase or reduce the soft excess.  %The recession of the inner edge of the disc due to Poynting-Robertson drag moves the inner reflecting zone farther away from the star, enhancing the soft excess. The disc inflation due to heating moves the upper disc surface closer to the neutron star, which increases the amount of received irradiation. 
Further studies on the dependence of the reflection spectrum characteristics to changes in the accretion disc during an X-ray burst may lead to probes of the disc geometry.%The reflection spectra calculated using the exact density structure predicted by the simulations are well approximated by constant density reflection models, especially near the burst peak. Thus, constant density models can continue to be used to model burst reflection spectra.
%We confirm that reflection is capable of producing the often observed soft excess. Constant density reflection models produce stronger emission lines and a lower soft excess, but the discrepancies are small enough to be still acceptable. A small soft excess may be an indicator of little disc recession.
\end{abstract}

\begin{keywords}
accretion, accretion discs -- radiative transfer -- stars: neutron -- X-rays: binaries --
X-rays: bursts
\end{keywords}

\section{Introduction } \label{sec:intro}

Reflection features are a common observable in the X-ray spectra of accreting compact objects of all mass scales \citep[e.g.,][]{Nandra1994,Xu2020,Szanecki2021}. The most prominent reflection features are the Fe K$\alpha$ line at $\approx6-7$ keV (e.g. \citealt{Fabian1991,Nandra1994,Fabian2000,ballantyne2002}; for a review see \citealt{Miller2007}) and the so-called Compton hump above 10 keV \citep[e.g.][]{Lightman1988, Fabian1991}. These features arise through reprocessed radiation. In binary systems, the reflection of radiation occurs primarily in the accretion disc  \citep[e.g.,][]{Cackett2010}. Due to the dependence on the reflector properties, reflection features are a powerful tool to study the accretion disc \citep[e.g.,][]{Ross1993,Ross2005,Garcia2011,Garcia2020}.

For neutron stars in low-mass X-ray binaries, the accretion disc is irradiated by a boundary layer \citep[e.g.,][]{Inogamov1999,Popham2001,Inogamov2010,Ding2021}, corona \citep[e.g.,][]{Galeev1979}, and, for some systems, by a Type I X-ray burst. Neutron stars in low-mass X-ray binaries accrete matter, predominantly hydrogen and helium, from a donor star via Roche-lobe overflow. The matter accumulates as an additional layer on the neutron star surface. This layer is subject to rising pressure and heats up until unstable nuclear burning sets off a conflagration detectable as an X-ray burst \citep[for a review see][]{Lewin1993,Galloway2021}. During a burst, the neutron star emits large amounts of soft X-ray radiation, which has detectable impacts on the neutron star environment \citep[for a review see][]{Degenaar2018}. The burst radiation Compton cools the corona and reduces its hard X-ray emission \citep[e.g.,][]{maccarone2003,Fernandez2020,Speicher2020}. As the burst emission supersedes the coronal and boundary layer emission as the dominant radiation intercepting the accretion disc, its reflection spectrum will change likewise. During bursts, observations record a soft excess \citep[e.g.,][]{zand2013,Keek2018,Bult2019,Chen2019}. In addition, Fe K and L lines may be detected during bursts \citep[e.g.,][]{Strohmayer2002,Degenaar2013,zand2013,Keek2014,Keek2017,Bult2019}.

%Time-resolved observations of Type I X-ray bursts are challenging due to their typically short duration, and are only now becoming more common . 
Time-resolved observations of Type I X-ray bursts are challenging due to their typically short duration of a few seconds \citep[e.g.][]{Keek2018,Bult2019}, and are only now becoming more common with observatories such as \textit{NICER} \citep{Gendreau2017}.
In the past, detailed time-resolved observations have been limited to longer bursts such as intermediate-duration and superbursts. \cite{Keek2017} found a 1 keV emission line weakening with decreasing burst flux during an intermediate-duration burst. Evolving reflection features with receding burst flux have also been observed during superbursts \citep[e.g.,][]{Strohmayer2002,Ballantyne2004apj,Keek2014,Keek2014ApJL}. \cite{Ballantyne2004apj} and \cite{Keek2014ApJL} used the changing reflection features to infer changes in the accretion disc structure, such as recession and inflation of the disc.

Recently, \cite{fragile_2020} simulated the time-dependent interaction of a thin accretion disc with a Type I X-ray burst during its rise. Their 2.5 s long simulations tracked the impact of a burst that peaked at 2.05 s with a peak luminosity of $3\times 10^{38}$ erg s\textsuperscript{-1}. In their simulations, the burst radiation heated the disc, raising the disc temperature by half an order of magnitude. As a result, the disc height doubled and its surface density and optical depth decreased by an order of magnitude. Moreover, increased Poynting-Robertson (PR) drag increased the accretion rate by up to an order of magnitude, moving the disc inner radius by several km \citep[see also][]{Walker1992,Worpel2013,Worpel2015}.

%The radiation due to the X-ray burst will imprint itself in the reflection spectrum. 

With facilities such as \textit{NICER}, time-resolved observations of Type I X-ray bursts are increasingly available \citep[e.g.][]{Keek2018}. Their analysis requires an accurate understanding of the reflection spectrum evolution of the disc. Here, we present a simulation-based prediction of the evolving reflection spectrum due to the rise of a burst. We employ the simulation data by \cite{fragile_2020} to track the reflection spectrum as the Type I X-ray burst rises to its peak. In section \ref{sec:calculations}, we outline the calculations. We present the results in section \ref{sec:results}, which we discuss in section \ref{sec:discussion}. We state our conclusions in section \ref{sec:conclusion}. 

\section{Calculations}\label{sec:calculations}
We aim to demonstrate time-dependent changes of the reflection spectrum due to a Type I X-ray burst using the simulation data by \cite{fragile_2020}. This section outlines the steps for processing the simulation data to obtain the needed input for our reflection spectra calculations.

\begin{figure}
    \centering
    \includegraphics[width = 0.4\textwidth]{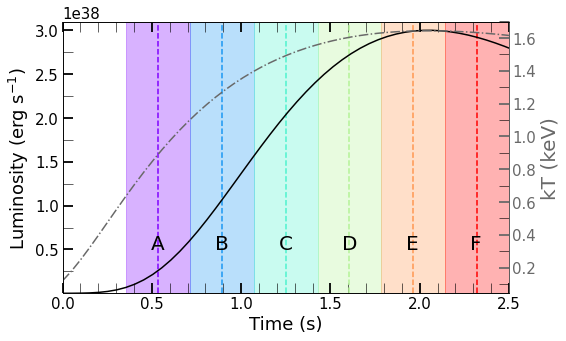}
    \caption{Time evolution of the burst luminosity and temperature from the \cite{fragile_2020} simulations. The burst luminosity (black solid line, equation \ref{eq:luminosity}) increases to a maximum of $3\times10^{38}$ erg s\textsuperscript{-1} at t=2.05 s. The shaded regions labeled A-F mark the six 0.357 s long time intervals we analyze, which start at t=0.357 s. Their midpoints used to calculate the luminosity per interval (equation \ref{eq:luminosity}) are marked by the dashed vertical lines. The burst blackbody temperature as seen from the disc is shown with the gray dot-dashed line (equation  \ref{eq:temperatureUpper}).}%radiates as a blackbody. Its temperature as seen from the disc is shown with the gray dot-dashed line 
    \label{fig:luminosityTime}
\end{figure}

In the simulation by \cite{fragile_2020}, an azimuthally symmetric disc in hydrostatic equilibrium and an initial mass accretion rate of $\dot{M} = 0.01L_{Edd}/c^2$, where $L_{Edd}$ is the Eddington luminosity, is illuminated by an X-ray burst. For our calculations, we adopt the same burst luminosity $L$ as \cite{fragile_2020}, which is given by \cite{Norris2005},
\begin{equation}
    L(t) = L_0 e^{2\left ( \tau_1/\tau_2 \right )^{1/2}} e^{-\frac{\tau_1}{t-t_s}-\frac{t-t_\mathrm{s}}{\tau_2}}.\label{eq:luminosity}
\end{equation}
In line with the simulations, $L_0=3\times10^{38}$ erg s\textsuperscript{-1}, $t_\mathrm{s} = -0.4$ s, $\tau_1=6$ s, and $\tau_2=1$ s \citep{fragile_2020}. Fig. \ref{fig:luminosityTime} depicts the burst luminosity evolution (black solid line). We focus on six 0.357 s long time intervals during the burst (A-F, shaded areas). The dashed vertical lines correspond to the interval midpoint, with which we calculate the luminosity in each time interval (equation \ref{eq:luminosity}).

We assume that the burst radiates as a blackbody and thus has a temperature $T_\mathrm{burst}$ as seen from the disc (Fig. \ref{fig:luminosityTime}, gray dot-dashed line),
\begin{equation}
    T_\mathrm{burst} = \left(\frac{L}{4\pi R^2_*\sigma}\right)^{1/4}\,\left(1+z_*\right)^{-1},\label{eq:temperatureUpper}
\end{equation}
where the gravitational redshift factor $(1+z_*)$ is \citep{Lewin1993}
\begin{equation}
    1+z_* = \left [ 1-\frac{2GM}{R_* c^2} \right ]^{-1/2},\label{eq:redshift}
\end{equation}
$\sigma$ is the Stefan-Boltzmann constant and $G$ the gravitational constant. Following \cite{fragile_2020}, the neutron star mass is assumed to be $M = 1.45$ M$_\odot$ and its radius is $R_*=10.7$ km.  

For each time interval, we process the time-averaged, two-dimensional simulation data by \cite{fragile_2020}. The data is organized in 384 half-spherical shells, subdivided into 384 cells each. Each cell is associated with a gas density and a radiation flux vector. With the density, we calculate the hydrogen number density $n_\mathrm{H}$ and the optical depth $\tau$ due to Thomson scattering across a cell, assuming a mean molecular weight $\mu = 0.615$.

The irradiation of the burst sets the boundaries of the reflecting region. Due to the fluid-like nature of the radiation in the simulation, the burst flux flows across the high density surface of the disc. We hence place the upper surface of our reflecting region where the average angle between the surface and radiation flux vectors is minimized. The average angles for our chosen surfaces are $\lesssim 3.4^\circ$. To check whether the selected surfaces capture the entire reflecting region, we integrate from the equatorial plane upwards, and find that small deviations at the location of the upper surface change the integrated $\tau$ due to Thomson scattering by $\lesssim 3.5\times10^{-5}$ for all time intervals. Thus, the chosen surfaces are located in the atmosphere above the bulk of the disc. 

\begin{figure*}
    \centering
    \includegraphics[width = 0.8\textwidth]{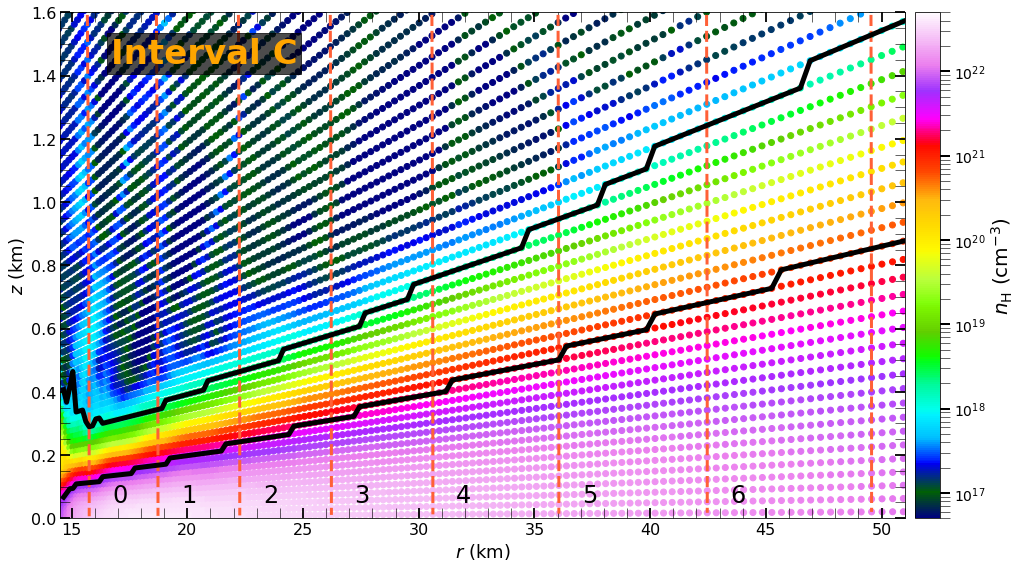}
    \caption{Gas density profile and reflecting region during time interval C. An upper and lower border (black solid lines) constrain the reflecting region. The upper border follows the disc atmosphere as defined by the procedure described in section \ref{sec:calculations}. The lower border is placed far enough to ensure the capture of the entire burst radiation and is at $\tau\approx 8$, where $\tau$ is the optical depth due to Thomson scattering integrated from the upper surface. The disc is divided into 7 radial zones (red dashed lines). Zone 0 is drained over time due to increased PR drag. Zones 1-6 are logarithmically spaced between $r\approx19$ km and $r\approx50$ km. They are of the same size for all time intervals. The gas inward of zone 0 is optically thin and does not contribute to any reflection features.}
    \label{fig:nHSlab}
\end{figure*}

Fig. \ref{fig:nHSlab} shows an example of the reflecting region for time interval C.
The upper surface (black solid line) traces the disc atmosphere, defined using the minimization technique described above. We set the lower boundary (black solid line) so that the region fully captures the burst and the resulting  temperature from X-ray heating  stabilizes towards the bottom. This occurs at $\tau\approx6$ for the time intervals A and B, $\tau\approx8$ for interval C, at $\tau\approx10$ for the intervals D and E, and at $\tau\approx12$ for interval F, if $\tau$ is integrated downwards starting at the upper surface. With the red dashed lines in Fig. \ref{fig:nHSlab}, we constrain and divide the reflecting region into radial zones to capture radial variations in the simulation. There is a decrease in resolution in the simulated disc beyond $r\approx64$ km and the outer regions have not reached equilibrium by the end of the simulation. Because the inner disc region is most resolved and contains the most significant structural disc changes, i.e. depletion of the innermost disc region due to PR drag and a change in height, we only consider the region within $r\lesssim 50$ km. 
%We therefore omit this region in our calculations.
As the burst proceeds, the inner accretion disc radius retracts, terminating the disc at $r\approx19$ km in time interval E. Within these bounds, 19 km $\lesssim r$ $\lesssim$ 50 km, we create six static logarithmically spaced radial zones (1-6). We account for the retracting inner accretion disc radius with an additional radial zone (0) that starts at an inner accretion disc radius, $r\approx15$ km, for intervals A-C. For interval D, we extend zone 1 to $\approx 18.4$ km instead of creating an additional zone 0. For interval F, the inner accretion disc radius has retracted further and the inner boundary of zone 1 is $r\approx19.2$ km. The gas inward of the innermost zone is optically thin and does not contribute to any reflection features. 

\begin{figure}
    \centering
    \includegraphics[width = 0.4 \textwidth]{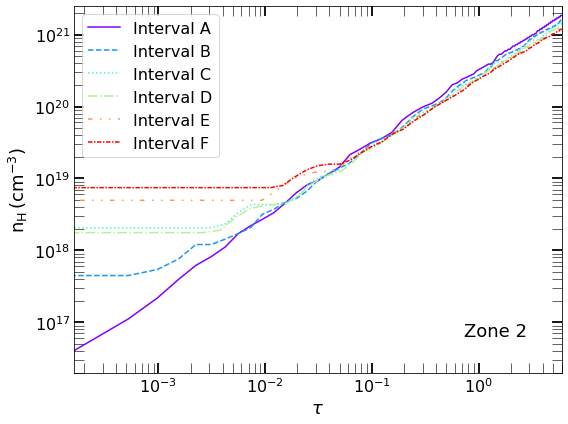}
    \caption{Vertical $n_\mathrm{H}$ profile for radial zone 2 for each time interval. The density (solid lines) increases within the reflecting region (Fig. \ref{fig:nHSlab}). The density increases for the first $\tau\lesssim5\times10^{-2}$ with time, but slightly decreases deeper within the reflection region. At $\tau\approx1$, $n_\mathrm{H}$ decreases by a factor of $\approx1.3$ from the onset of the burst (interval A) to its peak (interval E). Due to the small overall $n_\mathrm{H}$ variations, the time dependence of $n_\mathrm{H}$ will not be the main driver of the reflection spectrum evolution.}
    \label{fig:tauNH}
\end{figure}

For every radial zone in each time interval, we calculate a one-dimensional vertical $n_\mathrm{H}$ profile. For this profile, we average $n_\mathrm{H}$ across all shells within a radial zone along a $\tau$ grid, consisting of 110, logarithmically spaced entries that capture density variations within the reflecting region. The vertical $n_\mathrm{H}$ profile for the radial zone 2 is shown in Fig. \ref{fig:tauNH} for each time interval (solid lines). The greatest $n_\mathrm{H}$ variations over time occur only for the first $\tau\lesssim5\times10^{-2}$, over which $n_\mathrm{H}$ increases. The overall vertical $n_\mathrm{H}$ structure changes little, though. At $\tau\approx1$, where the bulk of the radiation-disc interaction will occur, $n_\mathrm{H}$ reduces by a factor of $\approx1.3$ from the onset of the burst (interval A) to its peak (interval E). In contrast to that, the average burst flux irradiating radial zone 2 increases by a factor of $\approx 7$ from interval A to interval E. Due to the small $n_\mathrm{H}$ decrease compared to the flux increase, we expect that the time-dependence of $n_\mathrm{H}$ has a relatively minor impact on the reflection spectra. Instead, the burst luminosity will be the main driver of the reflection spectrum evolution. This fact implies that constant density reflection models remain a good approximation to the emitted reflection spectrum (see Appendix \ref{subsec:ConstantDensity}).

The burst irradiates the upper surface of each radial zone with an incidence angle of $\cos\theta = 1/\sqrt{3}$, %. As already described, the radiation acts like a fluid and thus flows across the high density disc surface. While this property allows to determine the upper surface of the accretion disc, it does not translate into meaningful incidence angles. We therefore adopt the angle assumed in 
following the one-stream approximation of radiative transer \citep[][]{Rybicki1986}. In addition to the burst radiation, thermal radiation from the bulk of the disc is injected as diffuse radiation into the lower surface of each radial zone with temperature $T_\mathrm{th}$,
\begin{equation}
    T_\mathrm{th}= \left(\frac{\bar{F}_\mathrm{th}}{\sigma}\right)^{1/4},\label{eq:temperatureBottom}
\end{equation}
where $\bar{F}_\mathrm{th}$ is the corresponding average thermal flux, calculated from the simulations by averaging the flux magnitudes at the lower boundary of a radial zone in a time interval. 

%To compute the reflection spectrum in each radial zone, we assume that the burst radiation enters through the upper surface at an incidence angle $\cos\theta = 1/\sqrt{3}$ and a temperature given by equation (\ref{eq:temperatureUpper}) at the midpoint time of the respective time interval (Fig. \ref{fig:luminosityTime}, dashed lines). Thermal radiation from the bulk of the disc enters the lower surface of each radial zone with temperature $T_\mathrm{th}$,
%\begin{equation}
%     T_\mathrm{th}= \left(\frac{\bar{F}_\mathrm{th}}{\sigma}\right)^{1/4},\label{eq:temperatureBottom}
% \end{equation}
% where $\bar{F}_\mathrm{th}$ is the corresponding average thermal flux, calculated by averaging the flux magnitudes at the lower boundary of a radial zone in a time interval. 

The vertical $n_\mathrm{H}$ profiles, the average burst and thermal fluxes along with their temperatures are input into the code of \cite{Ballantyne2001}, which calculates the angle-averaged rest frame reflection spectrum for each radial zone and time interval. The burst irradiation propagates through the reflecting layer via a one-stream approximation \citep[][]{Ross1993}, while diffuse radiation is transferred via a Fokker-Planck equation \citep[][]{Ross1978}. The photons heat and cool the gas via Compton scattering, bremsstrahlung, the photoelectric effect, recombination, and three-body interactions \citep{Ross1979}. The code by \cite{Ballantyne2001} solves the equations of thermal and ionization equilibrium to determine the local temperature and fractional ionization in the $\tau$ grid defining the reflecting region \citep{Ross1993}.
%The progression of the burst radiation within the reflecting layer is calculated with a one-stream approximation \citep[][]{Ross1993}. The thermal radiation diffuses through the reflecting layer as described by the Fokker-Planck equation \citep[][]{Ross1978}. The reflecting layer is in hydrostatic equilibrium. 
The code accounts for the ionization stages C \textsc{v}–\textsc{vi}, N \textsc{vi}–\textsc{vii}, O \textsc{v}–\textsc{viii}, Mg \textsc{ix}–\textsc{xii}, Si \textsc{xi}–\textsc{xiv}, and Fe \textsc{xvi}–\textsc{xxvi}. We assume solar abundances. For each time interval, we integrate the reflection spectra of the radial zones by assuming that the upper surface forms the slant height of joint conical frustums, and we calculate their lateral areas for each radial zone.%The emitting area along the upper surface is calculated as the lateral areas of joint conical frustrums.

\section{Results}\label{sec:results}
\begin{figure*}
    \centering
    \includegraphics[width = 0.8
\textwidth]{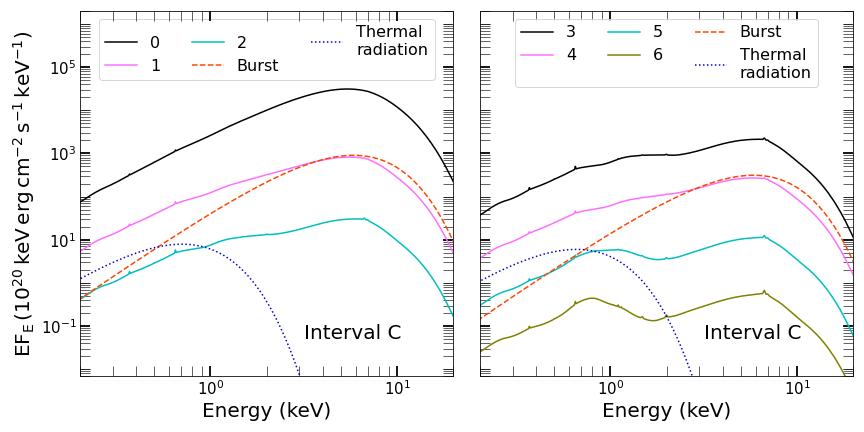}
    \caption{The solid lines show the reflection spectra of the radial zones, moving down, of interval C. For zone 1 (left panel) and 4 (right panel), we also plotted the burst (red dashed line) and the thermal emission (blue dotted line). The spectra were displaced by factors of 10 vertically for better visibility. The burst and thermal emission decrease with distance from the neutron star, causing the closest radial zones (left panel) to be the most ionized. Emission lines are pronounced in the less ionized outer radial zones (right panel), while closer radial zones exhibit more soft emission not attributed to the burst emission. Photoelectric absorption leads to an emission decrease at $\approx2$ keV, most noticeable for radial zones 5 and 6 (right panel).}
    \label{fig:individualSpectra}
\end{figure*}

Fig. \ref{fig:individualSpectra} shows the reflection spectra (solid lines) of all radial zones for time interval C. The spectra are displaced by factors of 10 vertically for better visibility. For radial zone 1 (left panel) and 4 (right panel), we also plot the burst emission (red dashed line) and the thermal radiation (blue dotted lines), which irradiate every radial zone from above and below respectively. Comparing the spectra of the radial zones closest to the neutron star (0-2, left panel) with the outer ones (3-6, right panel), it becomes apparent that reflection features increase in strength with distance from the neutron star. While the reflection spectra 0-2 bear the closest resemblance to the incident burst radiation (red dashed line) with few, weak emission lines, spectra 3-6 feature more, stronger emission lines due to carbon, oxygen, and iron. In addition, the outermost spectra (5,6) have a lower continuum at $\approx2$ keV due to photoelectric absorption. The difference in reflection characteristics stems from the radially decreasing density and irradiation. The density at $\tau\approx1$ decreases moving outwards by a factor of $\approx 3.6$ for interval C. However, the dominant factor is the decrease in burst flux. Across the radial zones, the average burst flux magnitudes drop by a factor of $\approx 45$ for time interval C. With decreasing burst flux, the degree of ionization decreases with distance from the neutron star, which enables stronger reflection features. Besides these features, a comparison of the reflected with the burst emission indicates the presence of a soft excess, soft emission not attributed to the burst emission (for further discussion, see section \ref{subsec:discGeometry}). 
%Reflection features increase in strength with distance from the neutron star. The difference in reflection features is mainly due to the difference in burst irradiation. $n_\mathrm{H}$ at $\tau\approx1$ decreases moving outwards by a factor of $\approx 3.6$ for interval C. However, across the radial zones the average burst flux magnitudes drop by a factor of $\approx 45$ for the same time interval. Radial zones closer to the neutron star (0-2, left panel) receive more burst (red dashed line, shown for zone 1 and 4) and thermal radiation (blue dotted line, shown for zone 1 and 4) compared to zones 3-6 (right panel) and are thus more ionized. The decrease in ionization as a function of distance can be seen by the strengthening of reflection features. Radial zones 3-6 show the strongest features, which are dominated by iron emission lines. Furthermore, the outermost radial zones (5,6) have a lower continuum at $\approx2$ keV due to photoelectric absorption. Comparing the reflected with the burst emission (plotted for the radial zones 1 and 4) indicates the presence of soft excess, soft emission not attributed to the burst emission (for further discussion, see section \ref{subsec:discGeometry}).%All radial zones display a soft excess, soft emission not attributed to the burst emission. Radial zones closer to the neutron star have a smaller soft excess, because the high burst flux promotes Compton cooling. At $\tau\approx1$, the cooling rate of Compton scattering is $\approx25$ times higher in zone 0 than in zone 6.

\begin{figure*}
    \centering
    \includegraphics[width = 0.9
\textwidth]{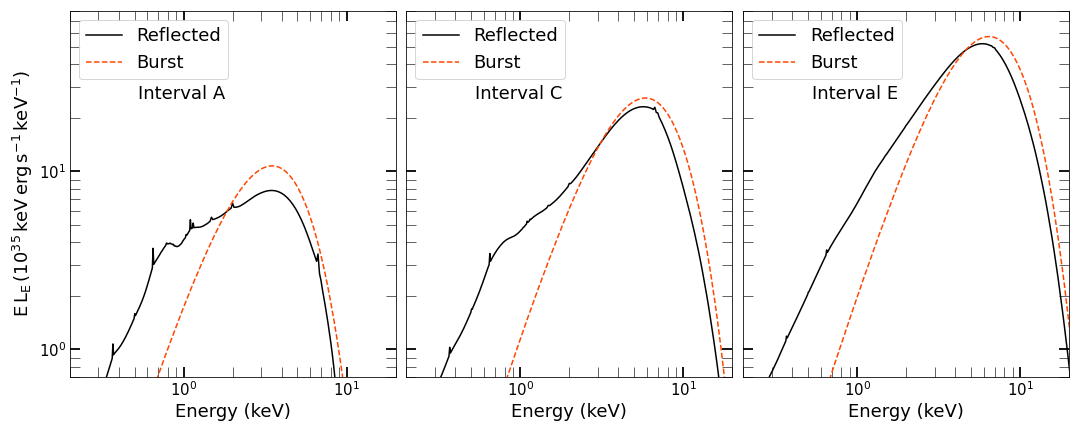}
    \caption{Radially integrated reflection spectra at different stages of the X-ray burst. The reflection spectra of the radial zones (Fig. \ref{fig:individualSpectra}) and the burst emission were integrated for each time interval. The burst (red dashed line) strengthens (equation \ref{eq:luminosity}, Fig. \ref{fig:luminosityTime}), increasingly ionizing and heating the accretion disc. The more intense irradiation gradually quenches any emission lines present in the reflection spectrum and increases the soft excess. The thermal radiation from the disc is not shown in the panels, but is included in the reprocessed emission.}
    \label{fig:finalSpectra}
\end{figure*}

The radially integrated reflection (black solid lines) and burst spectra (red dashed lines) for time intervals A, C and E are shown in Fig. \ref{fig:finalSpectra}. The thermal radiation is not shown, but is included in the reprocessed emission. As already anticipated in section \ref{sec:calculations}, the evolution of reflection features with time stems mainly from the evolution of the irradiation, which gains in intensity and energy. The burst irradiation increasingly ionizes and heats the accretion disc, increasing the amount of the soft excess. The advancing ionization weakens the emission lines. The emission lines are thus the strongest at the onset of the burst, as seen in the left panel. At $\approx1$ keV, Fe L lines are visible. At $\approx6.7$ keV, the spectrum features a recombination line of helium-like iron. We do not observe a Fe K$\alpha$ emission line at $\approx6.4$ keV, because the disc is highly ionized during all time intervals.
%Due to the burst emission not greatly surpassing the iron ionization energy, the recombination line of helium-like iron at $\approx6.7$ keV is weak. 
While Fe emission lines dominate the reflection spectrum, other ions such as C \textsc{vi} and O \textsc{viii} produce emission lines at $\approx 0.37$ keV and $\approx0.65$ keV. 
In time interval C (middle panel), the burst is brighter, ionizing the disc further and weakening emission lines, especially the Fe L lines at $\approx1$ keV. An emission line at $\approx7$ keV by hydrogen-like iron appears next to the weakened helium-like line at $\approx6.7$ keV. At the burst peak (right panel), the disc is almost completely ionized. Emission lines weaken severely or, like the lines at $\approx1$ keV, vanish altogether. 

\begin{figure}
    \centering
    \includegraphics[width = 0.4\textwidth]{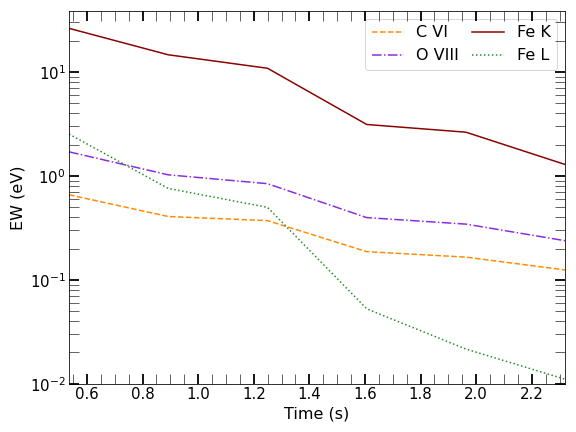}
    \caption{The EWs decrease over time due to the increasing ionization. The EWs were calculated following \cite{ballantyne2002} and only consider the reflection spectra. The Fe L lines (green dotted line) at $\approx1$ keV vanish during the burst rise; their combined EW decreases by $\gtrsim 99$\% towards the burst peak. The Fe K (dark red solid line) lines remain the strongest ones throughout the burst ascent but weaken by $\approx90$\% nevertheless. The C \textsc{vi} (orange dashed line) and O \textsc{viii} (purple dot-dashed line) lines are more stable during the burst. Their EWs decrease by $\approx75$\% and $\approx80$\% between interval A and E respectively.}
    \label{fig:ew_soft}
\end{figure}

The weakening of emission lines translates into decreasing equivalent widths (EW) over time (Fig. \ref{fig:ew_soft}), calculated following \cite{ballantyne2002}. When calculating the EWs, we neglect the diluting effect of the burst emission that is observed along with the reflection spectrum. We also ignore smearing due to disc rotation, which further decreases the EWs. On the other hand, we do not consider the disc beyond the 50 km radius, which will be less ionized, and therefore produce stronger emission lines. Longer simulations, where more of the simulated space reaches equilibrium, are needed to extend our calculations to a larger portion of the accretion disc. %Another factor yielding smaller EWs compared to observations could be the use of a varying density reflection model compared to a constant density one (section \ref{subsec:ConstantDensity}). 
The Fe K lines at $\approx6.7-7$ keV and the Fe L lines at $\approx1$ keV are spaced closely together respectively, so we report their respective cumulative EWs here. The Fe L lines (green dotted line) vanish during the burst ascent. Their cumulative EW decreases by $\gtrsim 99$\%. The Fe K emission lines (dark red solid line) are the strongest at all times, despite their EW dropping by $\approx 90$\% from the onset of the burst to its peak. C \textsc{vi} (orange dashed line) and O \textsc{viii} (purple dot-dashed line) form the most stable emission lines. Their EWs decrease by $\approx75$\% and $\approx80$\% at interval E compared to interval A.%\footnote{These predicted EWs omit the diluting effect of the burst spectrum that is observed along with the reflection spectrum. On the other hand, the EWs could be increased if disc radii beyond $50$ km were included in the predictions, as these parts of the disc will be less ionised than the inner region considered here. Longer simulations, where more of the simulated space reaches equilibrium, are needed to extend our calculations to a larger portion of the accretion disc.}. 

%Apart from disc ionization, the burst emission by itself and disc properties could affect the observed emission lines. When calculating the EWs, we neglected the diluting effect of the burst emission that is observed along with the reflection spectrum. Smearing due to disc rotation could complicate line detection further. On the other hand, we did not consider the disc beyond the 50 km radius, which will be less ionized, and therefore produce stronger emission lines. Longer simulations, where more of the simulated space reaches equilibrium, are needed to extend our calculations to a larger portion of the accretion disc. Another factor yielding smaller EWs compared to observations could be the use of a varying density reflection model compared to a constant density one (section \ref{subsec:ConstantDensity}).

While the EWs decrease as the burst rises, the amount of soft excess increases (Fig. \ref{fig:soft16}, black solid line). We calculated the percentage of soft excess below 3 keV  not attributed to burst irradiation as
\begin{equation}
    \mathcal{S} = \frac{\int^{E_\mathrm{h}}_{E_\mathrm{l}} \left[F_\mathrm{s}(E) - F_\mathrm{bb}(E)\right]\,\mathrm{d}E}{\int^{E_\mathrm{h}}_{E_\mathrm{l}} F_\mathrm{s}(E)\,\mathrm{d}E}\,\times100,\label{eq:softExcess}
\end{equation}
where $F_\mathrm{s}(E)$ is the integrated reflection spectra emission (Fig. \ref{fig:finalSpectra}, solid black line), and $F_\mathrm{bb}(E)$ the integrated burst emission at energy $E$ (Fig. \ref{fig:finalSpectra}, dashed red line). The energies $E_\mathrm{l}$ and $E_\mathrm{h}$ are $\approx$10\textsuperscript{-2} keV and $\approx$3 keV respectively. The soft excess $\mathcal{S}$ increases from $\approx $4\% from the burst onset to $\approx 38$\% at the burst peak. The trend in soft excess is due to Compton scattering and bremsstrahlung processes. The increasing burst luminosity increases the mean burst photon energy during the burst rise, increasing the Compton scattering rate. In addition, the increasingly energetic photons heat the disc, producing a rising number of bremsstrahlung photons. These photons undergo further Compton scattering in the hot, low-density disc surface on their way outwards. Some bremsstrahlung photons are thus upscattered and removed from the soft excess.

%. The increasingly energetic photons heat the disc. Due to the hotter disc, the inverse Compton scattering rate increases. In addition, the high temperatures of the disc produce a rising number of bremsstrahlung photons. These photons undergo further Compton scattering in the hot, low-density disc surface on their way outwards. Some bremsstrahlung photons are thus upscattered and removed from the soft excess.

\section{Discussion}\label{sec:discussion}
%Our calculations are the first self-consistent predictions of the reflection spectra evolution due to a Type I X-ray burst. We use novel simulation data by \cite{fragile_2020} and account for both the accretion disc and burst evolution to compute a changing reflection spectrum. We show that the strengthening X-ray burst increasingly ionizes the accretion disc, which weakens emission line strengths. The burst rise increases the Compton scattering and bremsstrahlung rate, which increases the amount of soft excess. Here, we will compare our results to observations and discuss the impact of accounting for the disc structure and its evolution.

\subsection{Implications for Observations of X-ray Bursts}\label{subsec:ComparisonObservations}

As X-ray bursts brighten over the course of seconds, time-resolved spectroscopy of the burst ascent is challenging. In this section, we therefore compare our results to studies focusing on the burst tail. While we expect that the reflection spectrum in the burst tail will approximately retrace the evolution shown in section \ref{sec:results} (Fig. \ref{fig:finalSpectra}), the burst tail will not be the exact reverse image of the burst rise. In the tail, the burst will continue to heat the disc and impact the disc geometry and thus its spectrum (section \ref{subsec:changingHeight}). Longer simulations of the disc-burst interaction lasting the entire burst are needed to predict the trends of observational properties more precisely.
Nevertheless, we believe that our qualitative results should apply to many objects. 
%While the reflection spectrum evolution in the burst tail will not exactly mirror the evolution in the burst rise, observational properties like the soft excess should follow the same trends. 
%Our analysis focused on the rest frame reflection spectrum only. By remaining in the rest frame, we do not observe shifting emission line centroids caused by gravitational redshift \citep[e.g.,][]{Fabian2000}, which have been used to constrain the inner accretion disc radius \citep[e.g.,][]{Ballantyne2004apj}. 

%The cooling times for bremsstrahlung and Compton scattering are much shorter than the timescales considered here. Observational properties like the soft excess should thus revert to their pre-burst values at the end of the burst.% A limitation of the reversal might be a major disruption of the accretion environment that does not recover over the course of the burst.    

A frequently observed feature is a soft excess increasing with burst luminosity and has either been treated as a byproduct of an increase in the mass accretion rate or due to reflection \citep[e.g.,][]{zand2013,Worpel2013,Worpel2015,Keek2018,Bult2019}. The observed soft excess persists even when the model of the burst emission is allowed to diverge from a blackbody at low energies \citep[][]{Fernandez2020}. Our calculations show that reflection can produce a strong soft excess. Similar to observations, the strength of the modeled soft excess rises with burst luminosity in the burst ascent (Fig. \ref{fig:soft16}). In fact, we find a soft excess starting from the onset of the burst, where the burst luminosity is still relatively low (Fig.\ref{fig:soft16}). Hence, a reflection-induced soft excess could be a standard feature of X-ray bursts regardless of their burst luminosity.

The burst also affects emission line formation. In our calculations, the rising disc ionization due to the rising burst luminosity weakens emission lines considerably. At the burst peak, the high ionization of the accretion disc renders many of the initially present emission lines undetectable. 
%many initially present emission lines. 
%the burst ionizes the accretion disc and significantly weakens emission lines over the course of the burst. Consequently, some or all emission lines may be unobservable during the peak of a bright burst. 
The difficulty in observing emission lines for a highly ionized accretion disc agrees with simulation results \citep[e.g.][]{Keek2016}, and measurements of spectra without emission lines during a bright burst  \citep[e.g.][]{Keek2018,Chen2019}. The burst blackbody temperature impacts emission line formation as well. Iron K lines, for instance, require burst photons with energies $\geq7.1$ keV \citep{Fabian2000}. In our calculations, the burst blackbody temperature peaked at $\approx1.6$ keV (eq.\ref{eq:temperatureUpper}, Fig.\ref{fig:luminosityTime}), consequently producing relatively few burst photons with energies $\geq7.1$ keV (see Fig.\ref{fig:finalSpectra}, orange dashed lines), which, in combination with the high disc ionization, contributed to the weakness of the Fe K lines in the calculated spectra. X-ray bursts can have stronger Fe K emission lines since observations yield burst blackbody temperatures up to $\approx3$ keV \citep[e.g.,][]{Hoffman1977,Ballantyne2004apj}, corresponding to more photons above the 7.1 keV threshold. However, we still expect these emission lines to follow the trend of our calculated lines, and to weaken with increasing ionization, regardless of the burst blackbody temperature.

%In our calculations, the relatively small amount of burst photons above that temperature threshold contributes, in addition to the high disc ionization, to the weakness of our Fe K lines. While other elements such as L, C, and O are less temperature-dependent and thus more common, their emission lines are also weaker. % Absent emission lines could lead to misleading conclusions about the composition of the accretion disc or the role of reflection during a burst. Therefore, we recommend emission lines studies to focus on the tail of the burst rather than on the burst peak, especially when the burst is luminous.

Despite different burst temperatures, our reflection calculations share similarities with the time-resolved studies of superbursts by \cite{Ballantyne2004apj}, and \cite{Keek2014ApJL}. \cite{Ballantyne2004apj} studied a superburst of 4U 1820--30, and \cite{Keek2014ApJL} examined a superburst of 4U 1636--536, both observed with \textit{RXTE}. As both bursts declined, the degree of ionization decreased by more than a factor of 10. In our calculations of the burst rise, we similarly observe that the disc ionization increases with burst strength. 
%In contrast to our calculations, the emission features observed by \cite{Strohmayer2002,Keek2014} strengthen during the burst. We measure a different emission feature behavior because the strong burst flux in our calculations ionizes our disc more over the course of the burst. %Due to less ionization during the burst compared to our calculations, the \cite{Strohmayer2002,Keek2014} observe a Fe K$\alpha$ line strengthening with increasing burst energy.  

%The emission lines weakened with decreasing ionization for both bursts \citep{Strohmayer2002,Keek2014ApJL}; the flux of the iron emission line at $\approx5.8-6.4$ keV present during the superburst of 4U 1820-30 dropped in the burst tail by a factor of $\approx4$ from its peak value \cite{Strohmayer2002}. That behavior is comparable to the Fe K line at $\approx7$ keV in our calculations, which only appeared for a higher degree of ionization. 
We expect more emission lines to form and strengthen during a burst with a smaller peak luminosity. As observed by \cite{Strohmayer2002} and \cite{Keek2014}, we anticipate a Fe K$\alpha$ line at $\approx6.4$ keV to form in a less ionized disc. With less ionization, we furthermore anticipate Fe L emission lines to strengthen. Previous burst observations detected an emission line at $\approx1$ keV \citep[e.g.][]{Degenaar2013,zand2013,Keek2017,Bult2019}. This line has often been attributed to iron and could be a combination of the emission lines we observe.  Other metals will form emission lines with lower ionization as well. With less ionization, we expect more emission line formation, especially at energies $\lesssim1$ keV as simulated by \cite{ballantyne2004}.  %A weaker burst will strengthen more Fe K lines, such as at $\approx6.4$ keV, which have already been observed during bursts \citep[e.g.][]{Strohmayer2002,Keek2014,Bult2019}. We furthermore anticipate Fe L emission lines to strengthen during a less luminous burst%That not more lines in our calculations are strengthened by increased ionization grounds in the high ionization of our accretion disc during the burst.

%\cite{zand2013} observed a X-ray burst from SAX J1808.4-3658 with \textit{Chandra} and RXTE. During the first 0-7 s, which includes the burst rise and most of its peak time, they detected an emission line at 1.04 keV, but not during the burst tail. During the burst tail, the registered burst luminosity is significantly lower than during the 0-7 s interval. We can thus assume that the burst photons were energetic enough to ionize the line producing ions, probably predominantly iron, only during the luminous phase of the burst. The link of the 1.04 keV emission line presence to a burst luminosity threshold is comparable to our detection of the emission line at $\sim7$ keV in interval C and E but not in interval A. The different emission lines featured during the burst evolution depend on the burst irradiation. With a lower burst luminosity, we will see emission lines at $\sim 1$ keV emerge at later burst stages and retain more of their strength. We will study the impact of a lower burst luminosity in future work.

\subsection{Impact of the Retreating Inner Accretion Disc Radius}\label{subsec:discGeometry}

\begin{figure}
    \centering
    \includegraphics[width = 0.4\textwidth]{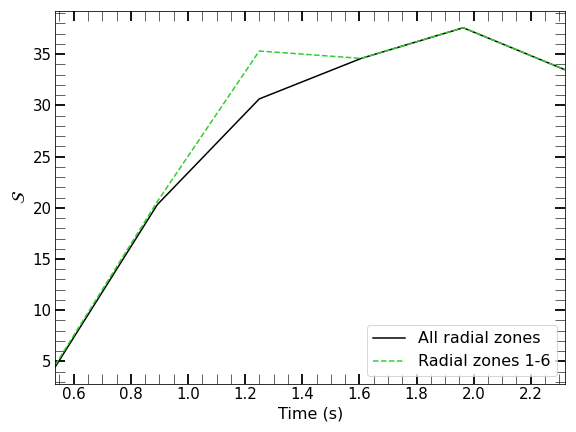}
    \caption{The percentage of soft excess $<$ 3 keV $\mathcal{S}$ (black solid line, equation \ref{eq:softExcess}). The soft excess increases during the burst rise. The increasing burst luminosity (equation \ref{eq:luminosity}) heats the disc, which strengthens Compton scattering and bremsstrahlung rates. Structural disc changes past the burst peak reduce the soft excess in time interval F (see also section \ref{subsec:changingHeight}). If the innermost radial zone 0 is neglected (green dashed line, section \ref{subsec:discGeometry}), $\mathcal{S}$ is larger compared to integrating over all radial zones (black solid line). The difference in $\mathcal{S}$ exists up to time interval C; time intervals D-F do not have a radial zone 0.}
    \label{fig:soft16}
\end{figure}

The burst interaction with the accretion disc affects the disc structure. PR drag drains the inner region of the accretion disc, causing the inner accretion disc radius to retreat \citep{fragile_2020}. Here, we discuss the impact of the retreating inner radius on the reflection spectrum.

We study the impact of the retreating accretion disc by comparing the soft excess of the total integrated reflection spectrum to the soft excess of the reflection spectrum only integrated over the fixed zones 1-6 (Fig. \ref{fig:soft16}, green dashed line). For the time intervals A-C, which have the extra zone 0, the soft excess of the spectrum integrated only over zones 1-6 (green dashed line) is higher than the soft excess of the total spectrum integrated over zones 0-6 (black solid line). The greatest difference in $\mathcal{S}$ occurs in interval C, where $\mathcal{S}\approx35\%$ for the spectrum of zones 1-6 compared to $\mathcal{S}\approx31\%$ for the total spectrum. 

The reason for the difference in $\mathcal{S}$ for time interval C lies in the high disc temperatures across the radial zones. In time interval C, the burst flux illuminating radial zone 0 is $\approx45$ times greater than for radial zone 6. The stronger burst flux in radial zone 0 heats the disc more compared to in zone 6. At $\tau\approx1$, the gas of radial zone 0 is $\approx2.2$ times hotter than of radial zone 6. The hotter gas increases cooling process rates. Compton cooling is more active in particular, $\approx25$ times higher for zone 0 than zone 6 at $\tau\approx1$ for interval C, and upscatters the produced bremsstrahlung photons out of the soft excess. High disc temperatures thus cause high cooling rates, which reduces the soft excess.    

Since the region closest to the neutron star receives the greatest amount of flux, it is subject to more disc heating. As disc heating affects the soft excess, the soft excess evolution could indicate the movement of the inner accretion disc. Little disc retreat amounts to a highly irradiated, hot zone close to the neutron star that emits less soft excess. Fig. \ref{fig:soft16} shows that the consequence of an inner, hot zone is a lower $\mathcal{S}$. A relatively small soft excess could therefore possibly point towards considerable disc heating and little disc retreat. 

\subsection{Impact of the Inflating Accretion Disc}\label{subsec:changingHeight}
\begin{figure*}
    \centering
    \includegraphics[width = 0.9 \textwidth]{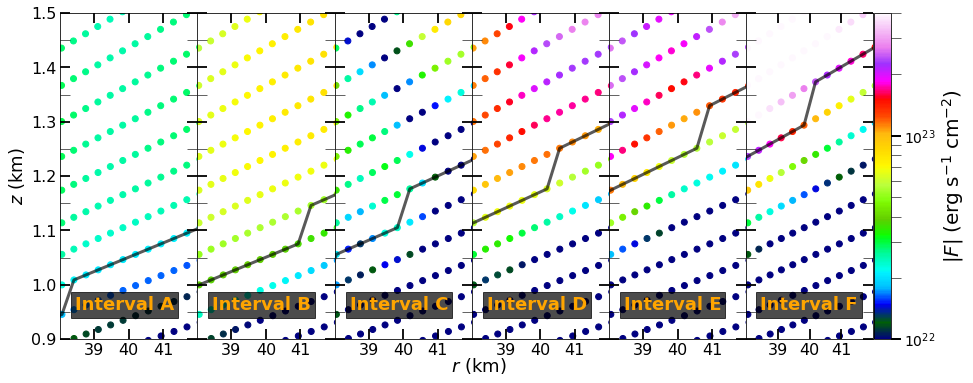}
    \caption{Segments of the upper surfaces at around $r\approx40$ km for time intervals A-F. The upper surface (black solid line) rises steadily due to disc heating, starting from time interval A (leftmost panel). Simultaneously, the burst flux magnitude (dots, color bar) increases. The rising disc surface remains highly irradiated throughout the burst, and leads to an $\approx 50$\% increase in the solid angle subtended by the reflecting surface. The increase in solid angle enhances the importance of reflection at the burst peak.}
    \label{fig:inflatingDisc}
\end{figure*}

The heating of the accretion disc due to the burst causes the accretion disc to inflate \citep{fragile_2020}, as Fig. \ref{fig:inflatingDisc} shows. The six panels feature a segment of the upper surface (black solid line) at around $r\approx40$ km for the time intervals A-F. The colored dots and the corresponding color bar show the burst flux magnitude per cell. The upper surface steadily rises, starting from time interval A (leftmost panel), even past the burst peak (interval F, rightmost panel). Simultaneously, the burst flux magnitudes at the upper surface in the shown segments increase. 

The flux evolution results in the upper surface of time interval F being the most irradiated, with an average flux $\approx1.7$ times higher than in interval E. Although interval F is just past the peak luminosity, interactions between the gas and radiation in the simulation lead  to regions of high flux despite a slightly lower luminosity injected at the inner boundary. As a result of this large flux, the disc in interval F is heated to the point where the soft excess extends beyond 3~keV (see also section \ref{subsec:discGeometry}), and $\mathcal{S}$ decreases to $\approx33\%$ (Fig. \ref{fig:soft16}). The high ionization further reduces the EWs of the emission lines in this interval (Fig. \ref{fig:ew_soft}). 

Disc inflation will influence the overall contribution of reflection in observations. As Fig. \ref{fig:inflatingDisc} shows, the rising upper disc surface begins to slowly turn towards the neutron star, which increases the solid angle subtended by the outer radial zones. For instance, the solid angle subtended by radial zone 6 is $\approx50\%$ larger in interval F compared to interval A. Disc regions further away from the neutron star will therefore be more irradiated and contribute more to the total reflected emission. Therefore, disc inflation will enhance the observed strength of highly ionized reflection close to the peak of the burst. As a result, despite the lack of emission lines, reflection is expected to provide a significant contribution to the total observed flux even past the peak of the burst.

%The burst luminosity just moved past its peak in interval F so that the simulated flux did not yet decrease but actually increased in parts of the simulated space compared to interval E. The region with increased flux encompasses the reflecting region, as a comparison of the two right most panels illustrates. 

%However, the increase of flux irradiating the upper surface in interval F is not solely due to an overall rise in burst flux, but also due to disc inflation, which elevates the upper surface to regions with stronger flux. For example, fixing the upper surface at its location of interval E would decrease the flux incident on the upper surface at $r\approx40$ km by a factor of $\approx0.3$ in interval F. Hence, disc inflation noticeably increases the amount of burst flux incident on the upper surface.%As already mentioned in section \ref{subsec:ComparisonObservations}, we do not expect the burst tail to be an exact reverse image of the burst rise, and longer simulations

% its upper surface to rise and increasingly face the neutron star. The inflation decreases the distance between the upper surface and the neutron star, increasing the flux on the upper surface. This effect may be illustrated with time interval F. While the burst luminosity declines in interval F, disc inflation continues. For example, the location of the disc at $r\approx50$ km moved from $z\approx1.5$ km in interval A to $z\approx2$ km in interval F. As a result of the closer distance to the neutron star, the disc receives a stronger burst flux in interval F compared to interval E. The burst flux magnitude irradiating the radial zones is on average $\approx1.7$ times higher in interval F than in interval E. 

%The high burst irradiation heats and ionizes the disc. The high amount of disc heating increases cooling processes (see also section \ref{subsec:discGeometry}), which decreases $\mathcal{S}$ to $\approx33\%$ in time interval F (Fig. \ref{fig:soft16}). The high ionization reduces the EWs of the emission lines further (Fig. \ref{fig:ew_soft}). %In contrast to previous time intervals, the high disc ionization erroneously suggests an increased burst luminosity in time interval F. Since disc inflation is partly responsible for this high ionization, the changing disc geometry weakens the link between burst luminosity and disc ionization. %Weak emission lines in connection to a small $\mathcal{S}$ could thus allude to significant disc inflation. 

\section{Conclusions}\label{sec:conclusion}
Type I X-ray burst spectra often show signs of reflection. Prominent reflection features are iron emission lines and a soft excess \citep[e.g.][]{Degenaar2013,zand2013,Keek2017,Keek2018,Bult2019,Chen2019}. Tracing the reflection features over time during the tail of superbursts first indicated substantial changes in the accretion disc structure \citep[][]{Ballantyne2004apj,Keek2014ApJL}. These changes were recently simulated by \cite{fragile_2020}. In this paper, we processed the simulation data of \cite{fragile_2020} to calculate the reflection spectra for six time intervals spanning the burst peak (section \ref{sec:calculations}). For each time interval, we divided the reflecting region into radial zones. For every zone, we determined vertical density profiles, the average burst and thermal fluxes as well as their temperatures. We computed the reflection spectra for each time interval and radial zone and finally integrated them.

The emission lines in the calculated spectra weaken due to the increased ionization over the course of the bursts rise (section \ref{sec:results}). While the calculated ionization evolution agrees with superburst observations (section \ref{subsec:ComparisonObservations}), the relatively low burst blackbody temperature combined with the high disc ionization in our calculations significantly limits the strength and types of emission lines we observe. Nevertheless, we expect observed emission lines to follow the calculated trend and weaken with increased ionization, regardless of the burst blackbody temperature.
%Our calculated reflection spectrum contains several emission lines that weaken due to increasing ionization over the course of the burst rise (section \ref{sec:results}). Our calculation results agree with the superburst observations regarding the disc ionization evolution (section \ref{subsec:ComparisonObservations}). The increasingly energetic burst photons heat the disc, which increases bremsstrahlung and Compton scattering rates. As a result, we measure a significant soft excess, increasing during the burst ascent with burst luminosity. 

We show that reflection is able to naturally produce a soft excess, which should be a standard feature of X-ray burst spectra. Our calculated soft excess is the product of Compton scattering and bremsstrahlung processes due to increasingly energetic photons heating the disc. The soft excess fraction evolves during the burst not only due to the unfolding burst but also because of the evolving disc structure; the drainage of the innermost disc region increases the soft excess (section \ref{subsec:discGeometry}), while disc inflation decreases the soft excess (section \ref{subsec:changingHeight}). The observed soft excess will be a combination of an excess due to reflection and an increase in the persistent spectrum due to an enhanced mass accretion rate \citep[][]{Worpel2013,Worpel2015}. Disentangling these two effects will allow a deeper understanding of the accretion disc geometry.

%A constant density model would change our results towards stronger emission lines and a smaller soft excess. However, the differences are small enough to be still acceptable, especially at the peak of the burst (section \ref{subsec:ConstantDensity}). 

Our work is a self-consistent prediction of a time-varying reflection spectrum and underlines the importance of studying dynamically resolved reflection spectra. The emission line evolution indicates the development of disc ionization and burst energy. Changes in soft excess could give insight into the disc geometry. To enhance the diagnostic capabilities of emission line features and the soft excess, their dependence on burst and disc characteristics must be known. Future work will probe the impact of burst and disc properties on the reflection spectrum.

\section*{Acknowledgements}
The authors thank P. Bult, T. G\"uver, and the amonymous referee for helpful comments that improved the manuscript. P.C.F. acknowledges support from National Science Foundation grants AST-1907850 and PHY-1748958.

\section*{Data Availability}
The data underlying this article will be shared on reasonable request to the corresponding author.

%may be used as a diagnostic tool for future dynamically resolved studies.
%This work presented the change of the reflection spectrum during a Type I X-ray burst. We showed that as the burst evolves, the accretion disc becomes increasingly ionized, suppressing emission line formation. As the burst strengthens and heats the disc, bremsstrahlung and Compton scattering are boosted, which translates into increasing soft excess over time. While the changes in the reflection spectrum mainly stem from the burst radiation, structural disc changes impact the spectrum as well as discussed in section \ref{subsec:discGeometry}. Future work will probe the impact of burst and disc properties on the reflection spectrum.

%Our studies may be used as a diagnostic tool for single time intervals or future more dynamically resolved studies. Constant density reflection models will be adequate tools for these studies, especially for tracking the amount of soft excess. 


\vspace{5mm}



\appendix
\section{Comparison to a Constant Density Reflection Model}\label{subsec:ConstantDensity}

Studies of X-ray bursts often rely on constant density reflection models \citep[e.g.][]{Ballantyne2004apj,Degenaar2013,zand2013,Keek2014,Keek2018,Bult2019}. However, the density in real accretion discs will vary vertically within the reflection region. Here, we explore how the assumption of constant density would influence the derived reflection characteristics.%discuss what impact a constant density reflection model would have on our results. 

To quantify the impact, we ran calculations with a vertically constant $n_\mathrm{H}$ structure for all time intervals. Because most interactions will occur at $\tau\approx1$, we selected the $n_\mathrm{H}$ values at this depth for each radial zone (Fig. \ref{fig:tauNH}). All other input values remained the same, and we computed the reflection spectra for each radial zone and time interval. As before, we integrated the reflection spectra of the radial zones to obtain the total reflection spectra per time interval.

\begin{figure*}
    \centering
    \includegraphics[width = 0.8 \textwidth]{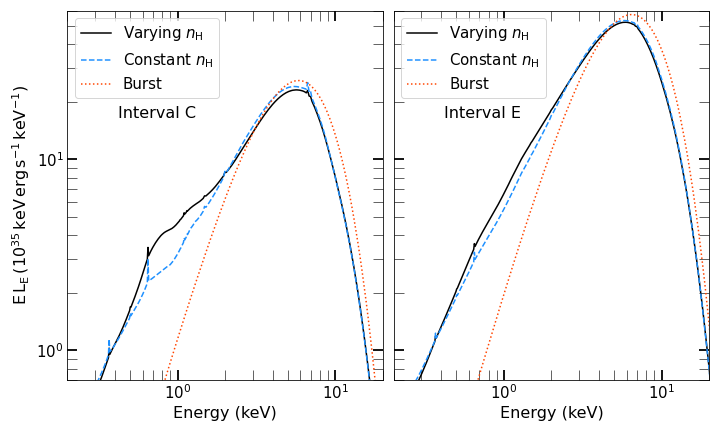}
    \caption{Same integrated reflection (black solid line) and burst spectra (orange dotted line) as in the middle and right panel of Fig. \ref{fig:finalSpectra}, but overlaid with integrated constant $n_\mathrm{H}$ reflection spectra (blue dashed line). For the constant $n_\mathrm{H}$ reflection spectra, $n_\mathrm{H}$ is vertically constant in each radial zone. The constant density spectra have a smaller soft excess and stronger emission lines because cooling processes and line formation occur in denser disc regions in the constant than in the varying density case. }
    \label{fig:finalSpectra_const}
\end{figure*}

\begin{figure}
    \centering
    \includegraphics[width = 0.4\textwidth]{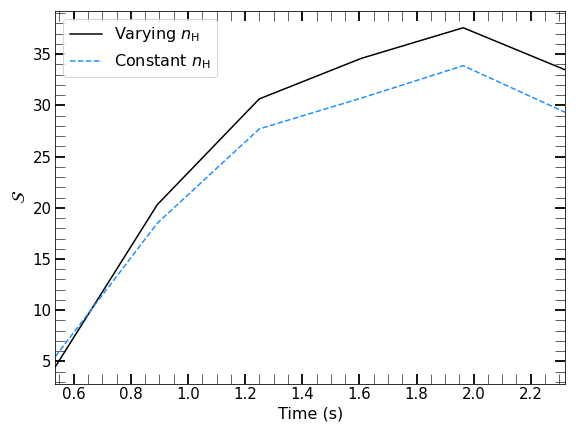}
    \caption{Same $\mathcal{S}$ as shown in Fig.\ref{fig:soft16}, but overlaid with the soft excess percentage $<$ 3 keV of the constant density reflection model. If $n_\mathrm{H}$ is varying vertically per radial zone (black solid line), $\mathcal{S}$ is up to $\lesssim1.1$ times greater compared to a constant density reflection model (blue dashed line).}
    \label{fig:soft_const}
\end{figure}

Fig. \ref{fig:finalSpectra_const} shows the difference of assuming a vertically constant $n_\mathrm{H}$ (blue dashed line) and the varying $n_\mathrm{H}$ profile taken from the simulation (black solid line) for time intervals C (left panel) and E (right panel). The spectra with the constant density assumption have a smaller soft excess at $\lesssim 2$ keV and stronger emission lines. These two differences result from a hot, low-density surface only present for the varying density profile.
%The high temperature in this surface raises the Compton cooling rate; for interval C at $\tau\approx10^{-3}$, the Compton cooling rate is $>200$ times higher across all radial zones for the varying than for the constant density profile. However, because the density in the hot surface is low for the varying density profile, few scatterings occur and few bremsstrahlung photons are scattered out the soft excess. 
The surface of the constant density profile is denser, which makes cooling processes, being two-body interactions, more efficient. %so that Compton scattering is more efficient in removing bremsstrahlung photons from the soft excess for the constant than for the varying density profile. 
Nevertheless, the lower soft continuum translates into a small $\mathcal{S}$ difference of $\approx28$\% with the constant density versus $\approx31$\%  with the varying density in interval C (Fig. \ref{fig:soft_const}, blue dashed versus black solid line).  

%This hot surface is subject to increased cooling processes through bremsstrahlung and most significantly through Compton scattering. For interval C at $\tau\approx10^{-3}$, the Compton cooling rate is $>200$ times higher across all radial zones for the varying than for the constant density profile. The cooling processes cool down the gas with the varying density profile quickly in a low-density region, though. 
%The hot surface promotes cooling through Compton scattering and bremsstrahlung. but also cools down the gas quickly. ence, with the constant density assumption the cooling rates through Compton scattering and bremsstrahlung are initially lower than for the varying density structure but higher
%In contrast, through the constant density assumption cooling processes engage with a higher density region.
%Since the gas with constant density is not rapidly cooled down, cooling processes are more active at $\tau\gtrsim1$ for the constant than for the varying density structure. Hence, cooling via Compton scattering and bremsstrahlung is more active for the constant than for the varying density structure for a larger, more dense part of the reflecting region, which causes the smaller soft excess for the constant density assumption. 

The emission lines with the constant density assumption are stronger because more metal ions are involved in line formation. For the varying density structure, ionization is concentrated in a thin, hot, low-density part of the disc. Line formation occurs at depths with higher metal ion density for the constant density assumption, which yields larger emission lines. For interval C, the O \textsc{viii} line is stronger by a factor of $\approx2.5$ for a constant density slab. On average, with a constant density the EWs are greater by a factor of $\approx2$ in interval C.

% At energies $\lesssim 2$ keV, the constant density spectra exhibit a smaller soft excess. The smaller soft excess is due to a higher disc temperature. In zone 6 at interval C, for example, the temperature at $\tau\approx1$ is $\approx2.3$ times greater if $n_\mathrm{H}$ is vertically constant. The higher temperature benefit cooling via Compton scattering and bremsstrahlung.   

% greater number of ions in the reflecting region also supports emission lines formation; Emission lines are stronger for a constant density reflection model, especially at low energies. 

The spectral shapes of the varying and the constant density models become more similar to each other as the burst rises, as shown by the spectra of interval E. % With the varying density structure, the density at small $\tau$ increases with time (Fig. \ref{fig:tauNH}). Therefore, the regions responsible for producing the soft excess and the emission lines for the different density profiles increasingly resemble each other. The density evolution of the varying density profile reduces the difference in emission line strength; the EWs for the constant density reflection model are on average 1.7 times higher for interval E. The O \textsc{viii} line has a $\approx2$ times greater EW if the density is constant. T
The discrepancy in the soft excess between the two models is minor for high luminosities. At the peak of the burst, $\mathcal{S}\approx34\%$ for the constant density model and $\approx38$\% for the varying one. Moreover, Fig. \ref{fig:soft_const} shows that $\mathcal{S}$ for a varying density model is greater by a factor of $\lesssim1.1$ than for a constant density model during the burst ascent. Constant density reflection models are thus adequate tools to study the soft excess. Their performance to study emission lines is also acceptable, especially at the peak of the burst. %However, the denser surface in the constant density reflection model may bias the derived ionization and abundance, which should be further investigated in the future.

%\section{Appendix information}

\bibliographystyle{mnras}
\bibliography{references}{}



% Don't change these lines
\bsp	% typesetting comment
\label{lastpage}
\end{document}